\documentclass[sigconf]{acmart}
\newcommand{\commentout}[1]{}
\usepackage{enumerate}
\usepackage{enumitem}
\usepackage{multirow}
\usepackage{booktabs}
\usepackage{graphicx}
\usepackage{caption}
\usepackage{color,soul}
\usepackage{subcaption}
\usepackage{fontawesome5}
\usepackage{tabularray}
\usepackage{hyperref}
\usepackage[toc,page]{appendix}
\usepackage{colortbl}
\usepackage{tcolorbox}
\usepackage{tikz}
\usepackage{tabularx}
\usepackage{makecell}

\definecolor{mygreen}{HTML}{20B77B}
\definecolor{myorange}{HTML}{FCCF4D}
\definecolor{mycyan}{HTML}{61D5C2}
\definecolor{myblue}{HTML}{90CBE4}
\definecolor{myred}{HTML}{EE754D}

\newcommand{\circleletter}[3][red]{%
  \tikz[baseline=(char.base)]{
    \node[shape=circle,draw=#1,fill=#1,text=white,inner sep=1pt] (char) {\textbf{#2}};
  }%
}

\newcommand{\cmt}[1]{}

\newcommand{\toolname}{\textsc{Aroma}}

\AtBeginDocument{%
  \providecommand\BibTeX{{%
    \normalfont B\kern-0.5em{\scshape i\kern-0.25em b}\kern-0.8em\TeX}}}

\copyrightyear{2025}
\acmYear{2025}
\setcopyright{cc}
\setcctype{by}
\acmConference[UIST '25]{The 38th Annual ACM Symposium on User Interface Software and Technology}{September 28-October 1, 2025}{Busan, Republic of Korea}
\acmBooktitle{The 38th Annual ACM Symposium on User Interface Software and Technology (UIST '25), September 28-October 1, 2025, Busan, Republic of Korea}\acmDOI{10.1145/3746059.3747650}
\acmISBN{979-8-4007-2037-6/2025/09}

\begin{document}

\title[]{\toolname: Mixed-Initiative AI Assistance for Non-Visual Cooking by Grounding Multimodal Information Between Reality and Videos}

\author{Zheng Ning}
\email{zning@nd.edu}
\affiliation{%
  \institution{University of Notre Dame}
  \city{Notre Dame}
  \state{Indiana}
  \country{USA}
}

\author{Leyang Li}
\email{lli27@nd.edu}
\affiliation{%
  \institution{University of Notre Dame}
  \city{Notre Dame}
  \state{Indiana}
  \country{USA}
}

\author{Daniel Killough}
\email{dkillough@wisc.edu}
\affiliation{%
  \institution{University of Wisconsin-Madison}
  \city{Madison}
  \state{Wisconsin}
  \country{USA}
}

\author{JooYoung Seo}
\email{jseo1005@illinois.edu}
\affiliation{%
  \institution{University of Illinois}
  \city{Urbana-Champaign}
  \state{Illinois}
  \country{USA}
}

\author{Patrick Carrington}
\email{pcarrington@cmu.edu}
\affiliation{%
  \institution{Carnegie Mellon University}
  \city{Pittsburgh}
  \state{Pennsylvania}
  \country{USA}}
  
\author{Yapeng Tian}
\email{yapeng.tian@utdallas.edu}
\affiliation{%
  \institution{University of Texas at Dallas}
  \city{Richardson}
  \state{Texas}
  \country{USA}}

\author{Yuhang Zhao}
\email{yuhang.zhao@cs.wisc.edu}
\affiliation{%
  \institution{University of Wisconsin-Madison}
  \city{Madison}
  \state{Wisconsin}
  \country{USA}}

\author{Franklin Mingzhe Li}
\email{mingzhe2@cs.cmu.edu}
\affiliation{%
  \institution{Carnegie Mellon University}
  \city{Pittsburgh}
  \state{Pennsylvania}
  \country{USA}}
  
\author{Toby Jia-Jun Li}
\email{toby.j.li@nd.edu}
\affiliation{%
  \institution{University of Notre Dame}
  \city{Notre Dame}
  \state{Indiana}
  \country{USA}}
\renewcommand{\shortauthors}{Ning et al.}

\begin{abstract} 
Videos offer rich audiovisual information that can support people in performing activities of daily living (ADLs), but they remain largely inaccessible to blind or low-vision (BLV) individuals. In cooking, BLV people often rely on \textit{non-visual} cues---such as touch, taste, and smell---to navigate their environment, making it difficult to follow the predominantly \textit{audiovisual} instructions found in video recipes. To address this problem, we introduce \toolname, an AI system that provides timely responses to the user based on real-time, context-aware assistance by integrating non-visual cues perceived by the user, a wearable camera feed, and video recipe content. \toolname\ uses a mixed-initiative approach: it responds to user requests while also proactively monitoring the video stream to offer timely alerts and guidance. This collaborative design leverages the complementary strengths of the user and AI system to align the physical environment with the video recipe, helping the user interpret their current state and make sense of the steps. We evaluated \toolname\ through a study with eight BLV participants and offered insights for designing interactive AI systems to support BLV individuals in performing ADLs. \looseness=-1
\end{abstract}



\begin{teaserfigure}
  \centering
  \includegraphics[width=\textwidth]{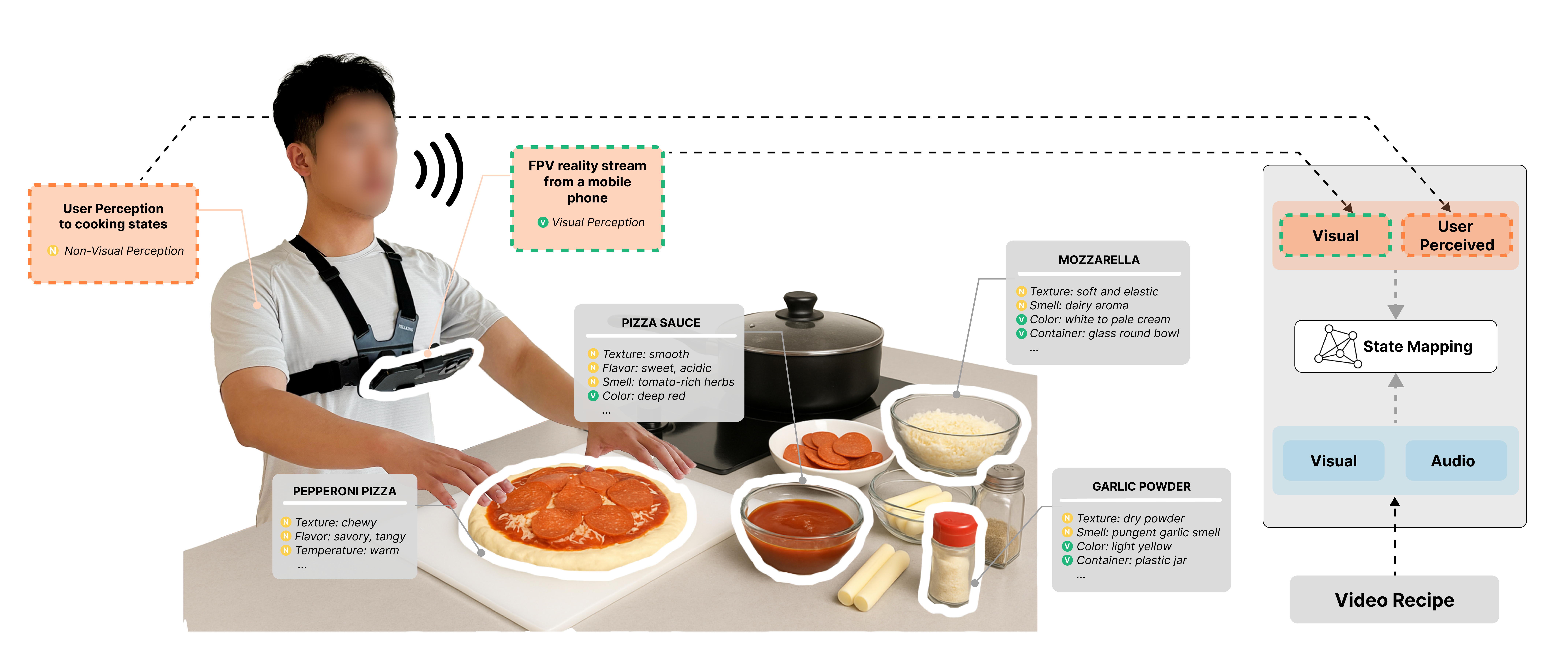}
  \caption{An illustration of how a blind or low-vision (BLV) user uses the \toolname~ in the kitchen. \toolname~ helps the user access video recipes by allowing them to communicate their perceived non-visual information (\circleletter[myorange]{N})) about the food, such as texture, smell, taste, etc., along with real-time visual information captured by a wearable camera (\circleletter[mygreen]{v})). The system then responds to their questions, referring to the knowledge from the video recipe during cooking. Meanwhile, \toolname\ proactively monitors the cooking process through the real-time video stream and raises alerts when specific criteria are met.}
  \Description{An illustration of a blind or low-vision user is shown cooking in a kitchen while interacting with the \toolname\ system. The user wears a smartphone mounted on a chest strap, with its camera directed at the food.}
  \label{fig: teaser}
\end{teaserfigure}

\maketitle

\section{Introduction}
\label{sec: intro}
Videos have become a critical resource for learning how to perform activities of daily living (ADLs)~\cite{li_freedom_2022,lin_identifying_2023}. For instance, platforms like YouTube have more than 500 hours of video content uploaded every minute~\cite{YouTubePressPage}, making it one of the most lively communities to share instructional content~\cite{lafreniere_community_2013}. As more instructions are shared exclusively in video format, videos have become an indispensable source for certain types of knowledge~\cite{merkt_learning_2011,bartolome_literature_2023}. Unlike single-modality formats such as text or static images, videos offer rich, multimodal instruction through visual demonstrations, spoken explanations, and audio cues. Despite their advantages, instructional videos for ADLs remain less accessible to blind or low-vision (BLV) individuals~\cite{li_recipe_2024}, as those videos are often designed with sighted viewers in mind. Specifically, in videos, instructors frequently rely on visual demonstrations without offering sufficient verbal detail, or they use vague references like “this” or “that” without context~\cite{li_recipe_2024}, making it difficult for BLV users to follow along.

Although efforts such as audio descriptions---narrated explanations of key visual elements, actions, and scene transitions~\cite{snyder_audio_2005,fryer_introduction_2016}---and video question-answering systems~\cite{castro_lifeqa_2020,yang_just_2021} aim to improve accessibility, instructional videos still pose challenges for BLV users attempting to perform tasks in the real world. One key issue is the mismatch between what is demonstrated in the video and what the user is experiencing in real-time. Specifically, BLV users primarily rely on non-visual sensory modalities, such as tactile feedback, smell, and sound, to track progress and make decisions during the activity~\cite{li_freedom_2022,li_contextual_2024}; however, these sensory modalities are frequently underrepresented or acknowledged in video content. As a result, BLV users often: (i) struggle to determine how their current state compares with what is shown in the video, and (ii) find it difficult to decide on the next appropriate steps based on their own sensory experiences.

Among the many activities demonstrated through instructional videos, cooking is particularly important. It is essential to support independence, health and emotional well-being~\cite{chang_recipescape_2018,min_survey_2019,li_recipe_2024}. Prior research shows that people with vision impairments are enthusiastic about video recipes, as they often include rich auditory cues (e.g., sizzling sounds) and are frequently produced by professional chefs~\cite{li_recipe_2024,li_non-visual_2021}. However, visually impaired cooks primarily consume these videos for entertainment or inspiration, rather than for direct task guidance, due to the lack of detailed visual descriptions~\cite{li_recipe_2024}. This disconnect highlights a fundamental challenge for BLV individuals — bridging the gap between their \textit{non-visual} sensory inputs (e.g., aroma, texture, or sound) and the instructional content primarily conveyed through \textit{visual} and \textit{auditory} modalities in video recipes. \looseness=-1

Prior research has offered valuable insights for designing AI systems that support BLV individuals in cooking by referencing video recipes~\cite{lin_identifying_2023,li_contextual_2024,li_recipe_2024}. However, a critical gap remains: enabling collaboration between users and AI agents that leverages their complementary perceptual and cognitive strengths to align real-time, multimodal information from the physical environment with instructional knowledge in the video.

To address this challenge, we introduce \toolname\footnote{\textsc{Aroma} is an acronym for \textbf{A}ugmenting \textbf{R}ecipe \textbf{O}rchestration with \textbf{M}ultimodal \textbf{A}ssistance}, an interactive AI system that enables real-time collaboration between a BLV user and a multimodal agent during cooking. Users can interact with the system to access procedural step information, verify the current state of cooking, and receive targeted guidance for correcting errors, etc., by directly querying the video recipe based on their own non-visual perceptions of the cooking process. Following a mixed-initiative design paradigm~\cite{horvitz_principles_1999}, the system also continuously monitors the cooking states through a wearable camera feed, proactively analyzing the scene and raising alerts when misalignments are detected.

We evaluated \toolname~ through a user study with eight BLV participants. Each user study session took place in the participant’s own kitchen or another preferred location by the participant. During the user study, each participant used the system to reproduce a dish of their choice from a list of three video recipes. Based on the results, we assess the system’s usability and identify key contextual challenges BLV users face when cooking with video-based instructions. We also propose design implications to inform the development of interactive AI systems aimed at improving video accessibility and supporting BLV individuals in performing ADLs.

To sum up, our paper presents the following contributions:
\begin{itemize}
    \item \toolname, an interactive mixed-initiative multimodal AI system that supports BLV users in following instructional cooking videos by leveraging the complementary perceptual and cognitive strengths of the user and the system to align information from the physical environment with instructional knowledge in the video.
    \item A user study with eight BLV participants to validate the usability and effectiveness of \toolname in realistic cooking tasks. 
    \item Design insights and findings for designing interactive AI systems that help BLV individuals access multimodal instructions and perform daily activities.
\end{itemize}
\section{Related Work}
\label{sec: relatedWork}
In this section, we review three key areas of related work that inform our research. First, we explore theoretical frameworks and assistive approaches designed to support individuals with sensory impairments in performing daily activities (Sec.~\ref{sec: rw_1}). Second, we discuss prior empirical findings and system work specifically addressing the challenges faced by BLV individuals in cooking (Sec.~\ref{sec: rw_2}). Lastly, we review research on AI models and interactive systems developed to enhance video content accessibility for BLV users (Sec.~\ref{sec: rw_3}). Throughout this review, we highlight how \toolname~ builds upon and extends these existing approaches by integrating real-time visual information with users' inherent non-visual perceptual capabilities in real-time mixed-initiative cognitive assistance for activities of daily living.

\subsection{Assisting People with Sensory Impairments in Performing Daily Activities}
\label{sec: rw_1}
Prior research has explored various approaches to support people with sensory impairments in performing daily activities, which can be categorized into two main areas: (i) facilitating cognitive processes to compensate for sensory input limitations such as visual, auditory, and olfactory~\cite{guo_statelens_2019,min_htike_augmented_2021,lucas_olfactory_2022,li2017braillesketch,kianpisheh2019face,jain_front_2023,ning_spica_2024,turkstra_assistive_2025}, and (ii) assisting users in overcoming physical barriers that hinder task execution~\cite{li_breaking_2023,li2021choose,li_it_2022,li_freedom_2022,li2025more}. Examples of the latter include designing Augmentative and Alternative Communication (AAC) for non-verbal users~\cite{valencia_aided_2021}, using eye-tracking devices for hands-free manipulation~\cite{higuch_can_2016}, designing accessible robots~\cite{dai_think_2024}, etc. Our work is more closely aligned with the first category, focusing on addressing cognitive challenges that arise when one or more perceptual modalities are impaired.

Key theoretical frameworks inform this space. The design principles of mixed-initiative user interfaces~\cite{horvitz_principles_1999} highlighted the importance of system-initiated actions in helping users detect and recover from errors. This is particularly crucial for individuals with sensory impairments, who may be unable to perceive such errors independently~\cite{lister_accessible_2020,prakash_improving_2024}. Similarly, multimodal disambiguation theory highlights the importance of integrating information across multiple modalities to more effectively interpret user intent~\cite{oviatt_mutual_1999}.

Building on those theories, prior research has applied assistive approaches across a variety of scenarios. To enhance navigation capabilities for BLV users, prior work has combined auditory and haptic cues to support spatial awareness~\cite{huppert_guidecopter_2021,schneider_dualpanto_2018,feng_haptics-based_2025}; Patil et al. explored using additional gestures on white canes to control the smart device of a BLV user~\cite{patil_gesturepod_2019}; Zhao et al. augmented the audiovisual information in a virtual reality (VR) scenario to enhance content understanding for low-vision users~\cite{zhao_seeingvr_2019}. For Deaf and Hard of Hearing (DHH) users, prior work has explored augmenting sound effects in VR~\cite{cao_supporting_2024}, adding extra indicators~\cite{li_soundvizvr_2022} and haptics~\cite{mirzaei_earvr_2020} to deliver a better experience for them; and captioning and visualizing non-speech sounds for a better video consumption experience~\cite{alonzo_beyond_2022}. Furthermore, the integration of olfactory feedback into VR environments has been explored to enhance user immersion and potentially aid those with olfactory impairments~\cite{liu_intelligent_2024,liu_soft_2023}.

The design rationales of \toolname~ are informed by the theories and insights discussed above. Specifically, cooking, as a naturally multimodal activity, requires individuals to integrate information from various sensory channels to understand and manage the process \cite{li_contextual_2024,li_non-visual_2021}. \toolname~ enables BLV users to communicate their non-visual sensory observations to the system and pairs this input with real-time visual analysis. This human-AI collaboration follows a mixed-initiative paradigm to bridge the gap between sensory perception and visual instruction, offering timely, context-aware support throughout the cooking process.

\subsection{Assisting BLV People in Cooking}
\label{sec: rw_2}
Cooking is an important daily activity that supports independence and improves the quality of life for blind or low-vision (BLV) individuals~\cite{chang_recipescape_2018,min_survey_2019,li_non-visual_2021}. Prior research has examined difficulties across various stages and design insights for assistive systems that support this task \cite{li_non-visual_2021}.

One major challenge lies in accessing and interpreting recipes. Text-based recipes are traditionally adopted because they are easy to follow through OCR~\cite{singh_survey_2012} and text-to-speech~\cite{openai_tts}; However, they often omit critical visual context. For instance, what is \textit{``cook until golden brown''} remains unclear~\cite{li_recipe_2024}. Video-based recipes, while rich in multi-modal information, introduce new difficulties, such as lacking structured navigation, insufficient verbalization of visual content, and increased cognitive load required to recall~\cite{li_recipe_2024,jiang_its_2024}. Although video recipes were reported as less accessible, the multimodal information was reported as entertaining and inspiring for people with vision impairments \cite{li_recipe_2024}. Recent research has also examined how BLV users engage with cooking instructions across modalities. For instance, Li et al. found that BLV cooks often prefer structured, chunked formats, tactile representations, and hands-free interaction mechanisms~\cite{li_recipe_2024}. Strategies like reordering steps, simplifying language, and annotating sensory checkpoints (e.g., smells, sounds) can enhance recipe usability.

Another significant hurdle is recognizing and interpreting real-time cooking states, such as ingredient readiness, food completion, or the location of utensils. Li et al~\cite{li_contextual_2024} conducted a contextual inquiry study, identifying eight classes of contextual information that BLV people actively seek. These include spatial layout, object status, and dynamic properties like temperature or completion, which serve important design rationales for \toolname. It has also pointed out that BLV users would develop intentional, embodied associations with objects (e.g., placing a spoon at a known angle) to ground information retrieval in spatial memory. These workarounds, however, are effective but fragile, especially in situations involving multitasking or shared kitchen environments. 

To address those issues, one relevant system is CookAR~\cite{lee_cookar_2024}, which augments the affordance of appliances and objects in cooking through a head-mounted AR system for low-vision users. Another similar system is OSCAR~\cite{li_oscar_2025,li2025exploring}, which provides context-aware feedback to the user when a task is completed by tracking object statuses. Specifically, OSCAR focuses on Step prediction in cooking, demonstrating that tracking object status changes in video significantly improves the accuracy, and step prediction is one of the important features in assisting non-visual cooking. In contrast, AROMA explores the dynamics of a real-time, human-AI partnership that focuses on assisting the cooking process as a whole, rather than the order of a particular step.

\toolname's voice-based model and mixed-initiative architecture are grounded in this body of research. The system is designed to align with BLV users' established preferences for interaction and feedback. For example, when describing a cooking step, the system provides a concise explanation that includes the step name, estimated duration, and expected outcome---an approach consistent with the preferred instruction style identified in prior work.

\subsection{Accessing Video Content for BLV Individuals}
\label{sec: rw_3}
Accessing video content remains a significant challenge for BLV individuals~\cite{liu_what_2021, liu_crossa11y_2022}. Prior research in this area can be broadly categorized into two areas: i) the support for sequential video consumption, where information is accessed in temporal order; ii) non-sequential information retrieval, where key information from the video is extracted based on user needs without requiring users to watch the entire video.

Audio descriptions (AD) are a primary method for enabling temporal video consumption by inserting narrated descriptions of visual content either inline (without pausing the video) or through extended descriptions (with pauses for additional narration)~\cite{aafaq2019video}. Prior research in this space has looked at automating the generation of ADs~\cite{wang_toward_2021} and new interaction paradigms such as layered audio descriptions that allow users to explore video content non-linearly while preserving the temporal flow~\cite{ning_spica_2024}.

When used in an information retrieval context, systems have been developed to help BLV users retrieve relevant information without watching the entire video. For example, Shortscribe~\cite{van_daele_making_2024} presents hierarchical summaries of short-form videos to quickly convey the gist. Another popular approach is to support question answering (QA) based on videos. For example, the AI community has made progress in Video Question Answering (VideoQA), where models generate natural language answers to queries grounded in video content directly~\cite{tapaswi_movieqa_2016, lei_less_2021, li_learning_2022}. 
Recent foundation models, trained on vast audiovisual corpora, have further pushed the boundaries of open-ended video understanding~\cite{gao_linvt_2024, google2024vertexai_video}.

The design of \toolname~ builds on these efforts by exploring how video recipes can serve as a rich instructional resource in real-world cooking scenarios, where the physical context often diverges from what is shown in the video. From a human-centered design perspective, \toolname~ also investigates how to leverage the non-visual perceptual strengths of BLV users, such as touch, smell, and sound, to support multi-modal interaction with video content, enabling users to cook effectively while accessing and aligning relevant information from the recipe video.
\section{\toolname~ System}
\label{sec:system}
\begin{figure}
    \centering
    \includegraphics[width=\linewidth]{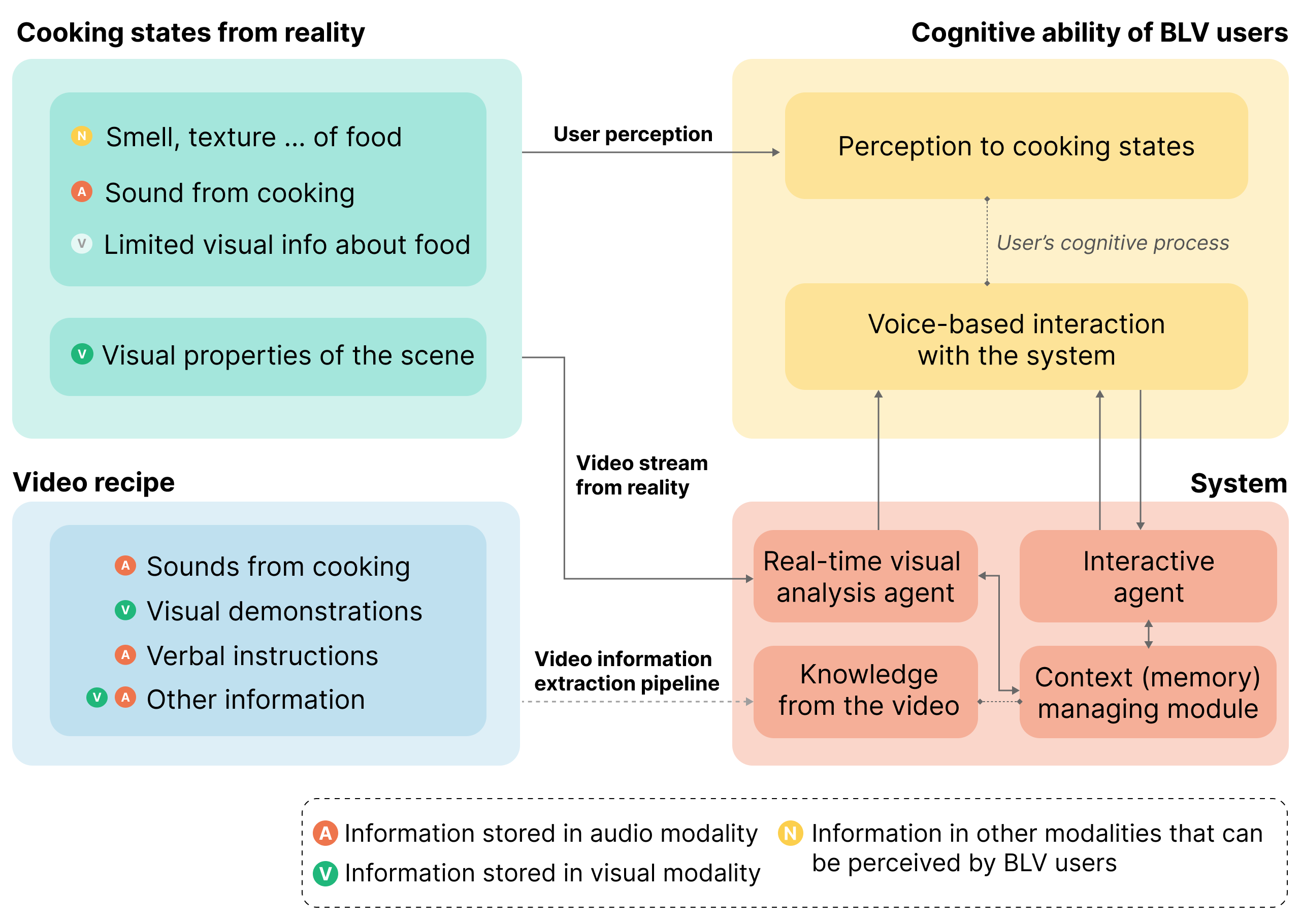}
    \caption{An illustration of the architecture of the system, and the corresponding input/output for each component. The system comprises four modules to extract knowledge from video recipes, analyze visual information from real-world cooking states, interact with the user through sound, and a module to manage history and agent context.}
    \Description{An illustration of the architecture of the system, and the corresponding input/output for each component. The system comprises four modules to extract knowledge from video recipes, analyze visual information from real-world cooking states, interact with the user through sound, and a module to manage history and agent context.}
    \label{fig: system_arch}
\end{figure}

We designed and implemented \toolname, a mixed-initiative system that couples BLV users’ non-visual cues with real-time visual information to support users in cooking. The system offers on-demand, conversational assistance and proactively detects errors during the cooking process by aligning information from the physical environment with instructional
knowledge in the video and provide timely and context-aware guidance to the user.

\subsection{Design Goals}
\label{sec: design_goals}
We identified the following design goals for \toolname, inspired by prior findings on non-visual cooking (detailed in Sec.~\ref{sec: rw_2}).

\begin{itemize}
    \item[DG1] \textbf{Provide both on-demand and proactive support.} The system should respond promptly to requests initiated by the user to address their immediate needs. This is particularly important given the cognitive demands of cooking and the difficulty of transferring knowledge from video recipes to real-world scenarios~\cite{woodworth_influence_1901,chandler_cognitive_1991}.  Additionally, the system should proactively monitor the cooking process to detect potential errors and offer corrective suggestions, recognizing that BLV users may not always be aware when a mistake occurs.

    \item[DG2] \textbf{Bridge the gap between real-world cooking states and video content.} The system should fuse the user’s non-visual perception knowledge \textit{(``I can feel the dough is sticky.)''} with real-time visual analysis \textit{(``The dough is too thin compared to the recipe’s demonstration.'')} to align the user’s actual cooking state with the intended recipe steps. In addition, the system should flexibly align real-world sensory input with video instructions, rather than enforcing rigid mappings. 
    To be effective, it must also interpret and adapt to the user’s context by handling ambiguities and offering disambiguation strategies when needed.

    \item[DG3] \textbf{Support flexible and accessible interaction with video recipes.}  Referencing video recipes during cooking can impose a high cognitive load due to their length and multi-modal nature~\cite{li_non-visual_2021,li_recipe_2024}. The system should accommodate diverse user needs by offering multiple forms of support, such as concise step-by-step guidance, detailed explanations, and the ability to jump directly to relevant video segments.
\end{itemize}

\subsection{Example Usage Scenario}
This section presents an example usage scenario in which Jane, a congenitally blind user, uses \toolname~ to prepare Spaghetti Bolognese. Each system feature mentioned corresponds to those described in Sec.~\ref{sec: key_features}, and is labeled as Ft.\{ID\}.

\vspace{0.5em}
\noindent\textbf{Preparation} Jane starts by watching a video on the recipe and listens to it from beginning to end. While this gives her a general idea of the steps, she cannot remember everything, and many details are unclear. To get more help, she opens ~\toolname~ and loads the video. The system first analyzes the pre-recorded instructional video; it then uses Jane’s mobile phone as a first-person camera to stream the kitchen environment in real time while capturing her voice commands for processing.

\vspace{0.5em}
\noindent\textbf{Step 1: Boiling Pasta} Following the tutorial, Jane fills a pot with water and places it on the stove to boil.  As Jane fills the pot and brings the water to a boil, the real-time monitoring agent aligns and compares her actions with the reference from the corresponding tutorial video. It detects that Jane has forgotten to add salt. In response, the system proactively alerts Jane with: \emph{``It appears you have not added salt to your boiling water. Adding salt enhances the flavor of the pasta.''} (Ft.2).

\vspace{0.5em}
\noindent\textbf{Step 2: Sautéing Onions and Garlic} Jane recalls that her next step is to sauté chopped onions and garlic in olive oil. While preparing the ingredients, Jane feels uncertain about the knife technique, so she asks, \emph{``How do I properly chop onions?''}. The system responds with: \emph{``Onion should be chopped into small chunks''}, which was retrieved and synthesized from the video content (Ft.1). Jane finds the initial response too general, so she tries chopping the onion for a few samples, using her sense of touch to judge the thickness. She then follows up with the system, showing a sample and asking, \emph{``I'm chopping my onion this thin, is this correct?''} After a brief pause, the system replies, \emph{``According to the video recipe, the onions are chopped into thicker square slices. For making sauce, there's no need to chop them finely.''} (Ft.1). Jane is satisfied with the answer and decides to continue cooking.

\vspace{0.5em}
\noindent\textbf{Step 3: Cooking the Sauce}
After chopping the onions, Jane forgets what her next step is, so she asks: \emph{``What's my next step?''}. The system automatically takes the history of her previous actions into consideration, retrieves the content from the video, and responds to her with: \emph{``Next, you should prepare the sauce. Heat oil in a pan, sauté the onions and garlic, and then incorporate tomato sauce and herbs.''} (Ft.1). Jane feels uncertain about this response, so she says \emph{``play the video recipe''} to ask the system to replay the relevant segments about \textit{chopping onions} from the original video recipe. Jane can also control the clip by saying \emph{``pause''} or related commands (Ft.3).

\vspace{0.5em}
\noindent\textbf{Step 4: Combining Pasta and Sauce}
In the final step, Jane needs to drain the pasta and mix it with the sauce before serving with grated Parmesan cheese. Here, she wants to reflect on the previous step, so she asks, \emph{``Did I drain the pasta already or is it still in the water?''} After a short thinking process, the system replies based on the memory of the process (Ft.1).

\subsection{System Overview}
\label{sec: system_arch}

\begin{figure*}[!htb]
    \centering
    \includegraphics[width=\linewidth]{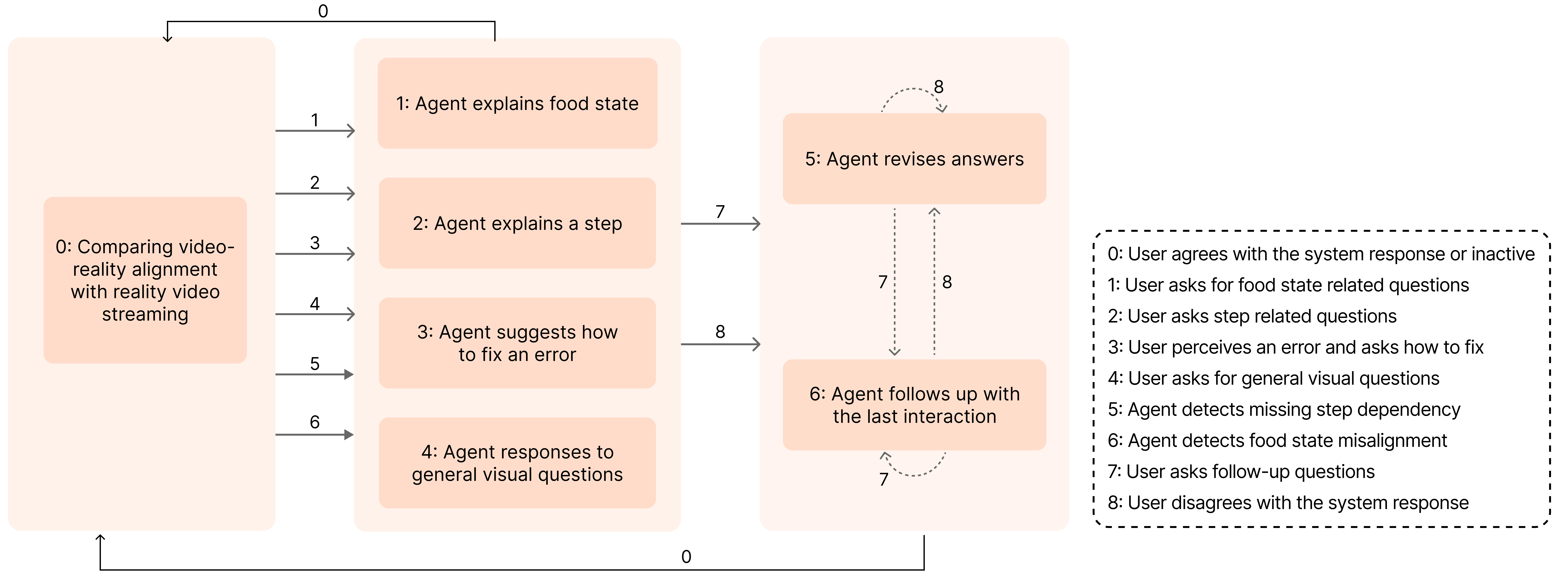}
    \caption{The state-machine-like framework in \toolname\ to handle various user queries and reality visual scene changes. The system starts at the initial state (state 0), and transitions between different states are determined by various events. Each event is decided by the current visual scene and the user's query (if available) by an LLM. When transitioning to a new state, an LLM generates a response using a predefined prompt tailored to that state. The response is subsequently transformed into audio output through a text-to-speech (TTS) service and delivered to the user.}
    \Description{The state-machine framework in \toolname: there are in total 9 states and 10 events. Each state determines the template of the system response. Transitions between different states (events) are determined by the visual scene and the user's query (if valid). The state machine automatically returns to the original state when the user is satisfied or when the system receives no input for a certain period. }
    \label{fig: state_machine}
\end{figure*}

\toolname~ operates through a mixed-initiative architecture. Rather than relying solely on user prompts or system-driven guidance, \toolname~ continuously integrates: \textit{non-visual perceptions} (see Fig.~\ref{fig: system_arch} {\textcolor[HTML]{FCCF4D}{\rule{0.6em}{0.6em}}}): the BLV user’s own sensory cues, such as smell, touch, and taste, conveyed verbally (e.g., describing the food texture or tasting for saltiness); \textit{visual information} (Fig.~\ref{fig: system_arch} {\textcolor[HTML]{20B77B}{\rule{0.6em}{0.6em}}} \circleletter[mygreen]{v})): real-time analysis of cooking states and ingredients from a wearable first-person camera (shown in Fig.~\ref{fig: teaser}); and \textit{video recipe knowledge} (Fig.~\ref{fig: system_arch} {\textcolor[HTML]{61D5C2}{\rule{0.6em}{0.6em}}}): information extracted and structured from the given video recipe.

There are four primary components (see Fig.~\ref{fig: system_arch} {\textcolor[HTML]{EE754D}{\rule{0.6em}{0.6em}}}) to coordinate reality information input in different modalities and process user interaction. Specifically, knowledge from the corresponding video recipe is stored in an accessible JSON form from a video information extraction pipeline. In parallel, a real-time visual analysis agent (implemented with a visual LLM, details in Sec.~\ref{sec: implementation}) streams the reality visual information into the system and monitors the cooking states by aligning the visual information with the video content. An interactive agent implemented based on another LLM handles the users' requests based on their perceived non-visual information, visual information from the visual agent, and the context. Video recipe knowledge, the periodically analyzed kitchen visual information, along with all user–agent conversational data, is stored as memory and retrieved by the context management module to provide adaptive context for each system response.

To coordinate user requests with evolving kitchen scenes, we model interaction as a deterministic state machine containing nine events and seven states drawn from prior studies \cite{li_non-visual_2021,li_recipe_2024,li_contextual_2024}. Fig.~\ref{fig: state_machine} illustrates this framework. Each event is identified by the LLM based on the current visual scene and the user's voice request. For agent-initiated events (events 5 and 6), there was no need for user voice information. Every state links to a prompt template that guides the backend LLM. When the system enters a new state, it generates a response based on the current sensory input, video recipe knowledge, and the session context.

\subsection{Key Features}
\label{sec: key_features}
Following the mixed-initiative interaction paradigm, \toolname~ implements the following features to fulfill the design goals outlined in Sec.~\ref{sec: design_goals}. 


\subsubsection{Ft.1: Contextualized Responses Grounded in Visual and Non-Visual Perception}
\label{sec: ft_user_init}
Since cooking involves multi-modal information (as shown in Fig.~\ref{fig: teaser}), where both visual and non-visual inputs are important to decide a cooking state, \toolname\ integrates visual data --- captured through a wearable camera --- and non-visual information verbalized by the user to represent the cooking state. This combined perceptual input is then grounded in the instructional content derived from the video recipe, as illustrated in Fig.~\ref{fig: teaser}.

Upon receiving a user query, the system immediately leverages the query that integrates visual and non-visual information to determine the current \textit{event}, defined in the state machine framework (see Fig.~\ref{fig: state_machine}), and the system transitions into one of the following four response categories:

\begin{itemize}
    \item \textbf{Food State Responses} (state 1 in Fig.~\ref{fig: state_machine}): inquiries about the current condition of the food. Triggered by event 1 from state 0.
    \item \textbf{Step-Related Responses} (state 2 in Fig.~\ref{fig: state_machine}) --- Clarifications regarding specific steps in the cooking process. Triggered by event 2 from state 0.
    \item \textbf{Problem-Solving Responses} (state 3 in Fig.~\ref{fig: state_machine}) --- Assistance when encountering issues or uncertainties. Triggered by events 3, 5, and 6 from state 0.
    \item \textbf{General Visual Questions Responses} (state 4 in Fig.~\ref{fig: state_machine}) --- Broader inquiries related to visual guidance on cooking procedure or recipe. Triggered by event 4 from state 0.
\end{itemize} 

Responses are designed to deliver concise and critical information relevant to the user's immediate needs. For example, if a user inquires, \textit{``Is the chicken cooked through?''} The system provides a targeted response as: \textit{``The chicken is lightly browned externally but requires additional cooking, as the internal temperature has not reached 165°F.''} --- deliberately omitting extraneous details to minimize cognitive load.

If the query is either a follow up for more details (event 7 in Fig.~\ref{fig: state_machine}) or the previous error wrong (event 8 in Fig.~\ref{fig: state_machine}) --- particularly because a user is unsatisfied with a default response due to vagueness, ambiguity, or limitations in model interpretation --- the state will transit to the corresponding one (state 6 or state 5 in Fig.~\ref{fig: state_machine}, respectively) where the backend LLM of the agent is prompted to pay special attention to the additional information provided by the user. \looseness=-1

The system adapts dynamically to the user’s non-visual perceptual input, such as information obtained through touch, sound, or smell, and incorporates that context into more grounded responses. This feature leverages the language model's capacity to synthesize multi-modal information and maintain context over multi-turn interactions.

\subsubsection{Ft.2: Proactive Monitoring of the Cooking Process}
\label{sec: ft_proactively_monitoring}
Previous research~\cite{li_contextual_2024} highlights the increased cognitive load BLV users face when accessing recipe instructions during cooking, which often prevents them from noticing errors during the cooking process. To address this challenge, \toolname~ adopts a mixed-initiative interaction model~\cite{horvitz_principles_1999}, blending user-initiated requests with system-initiated assistance to reduce friction and offload cognitive effort.

Specifically, \toolname~ leverages a video analysis agent that continuously monitors the cooking environment through a first-person camera (see Fig.~\ref{fig: teaser}). It performs two key operations every two seconds: (i) it generates objective observations of the current scene, and (ii) it compares those observations with reference knowledge extracted from the video recipe to make judgments.

Specifically, the agent is instructed to observe and describe:

\begin{itemize} 
    \item The specific cooking action being performed 
    \item The corresponding recipe step 
    \item The visible food items, ingredients, and kitchenware 
    \item Any identifiable cooking-related sounds 
\end{itemize}

Based on these observations, the agent determines:

\begin{itemize} 
    \item Whether the observed activity is relevant to the recipe 
    \item If relevant, whether the step is being executed correctly 
    \item Whether any required steps have been missed 
    \item Whether the user has advanced to a new step 
\end{itemize}

When the agent detects a deviation, such as a missed or incorrectly performed step, it alerts the user and provides corrective instructions. These situations map to event 5 and event 6 in Fig.~\ref{fig: state_machine} in the state machine, prompting a transition to a new interaction state (state 3 in Fig.~\ref{fig: state_machine}), which in turn triggers the appropriate system response.

\subsubsection{Ft.3: Accessing Video Segments and Memory Stored}
\label{sec: ft_access_offline}
To support DG3, \toolname\ enables users to access not only information from the current scene but also segments from the original instructional video, and memory of previous user-agent interactions and visual information stored by the real-time visual analysis agent every given time interval (2 seconds in our setting).

To access video segments, users simply make a voice request e.g. \textit{``Replay the part that tells me what ingredients I should prepare''}, and the backend LLM will interpret the semantic meaning of the request and play the segmented parts automatically to the user. Noticeably, the system automatically retrieves the relevant video segments from the video for the response it generates from Ft.1 (Sec.~\ref{sec: ft_user_init}) or Ft.2 (Sec.~\ref{sec: ft_proactively_monitoring}). This allows the users to quickly refer back to the video recipe to find \textit{evidence} if needed. To do this, users can issue commands like “play” or “pause” after receiving a response. This feature is designed to complement the conversational guidance and reduce cognitive effort. The design rationale for it is grounded on prior research, which highlights the effectiveness of AI-provided concrete examples in strengthening user trust~\cite{cai_effects_2019}. In the context of cooking, it is also a common practice for users to navigate to the appropriate parts from the original recipe~\cite{li_contextual_2024}.

To further support DG1 and DG2 and to address the practical challenges of managing multi-step cooking tasks, the system also allows users to retrieve information from earlier stages of the session. \toolname\ maintains a comprehensive record of both the conversational history and the automatic visual analysis results every 2 seconds. Users can access this information by making a retrieval-related request, such as \textit{``Did I already add the garlic?''}, and the system will retrieve the information from the context (memory) managing module (Fig.~\ref{fig: system_arch} {\textcolor[HTML]{EE754D}{\rule{0.6em}{0.6em}}}) and play it to the user automatically, enabling users to reflect on or resume prior steps without guesswork\looseness=-1.

\subsection{Implementation Details}
\label{sec: implementation}
The system is implemented using Next.js\footnote{\url{https://nextjs.org/}}. We use OpenAI Real-time API\footnote{\url{https://platform.openai.com/docs/guides/realtime}} to continuously transcribe users' voice commands. The real-time visual analysis agent is achieved by periodically making the same request to the Multimodal Live API from Gemini\footnote{\url{https://ai.google.dev/gemini-api/docs/live}}. In addition, we used GPT-4o-mini\footnote{\url{https://platform.openai.com/docs/models/gpt-4o-mini}} to generate user-initiated responses (see Sec.~\ref{sec: ft_user_init}) and extract relevant clips (Sec.~\ref{sec: ft_access_offline}) from the original video at the sentence level. We use OpenAI Text-to-Speech API\footnote{\url{https://platform.openai.com/docs/guides/text-to-speech}} to convert text-based responses into audio. 

To extract audio and visual information from video recipes, we use a set of AI models to obtain text-based descriptions from each modality separately. We first separate the vocal part from the video --- typically containing spoken instructions --- and store the transcribed text. We then parse the transcript at the sentence level and use PySceneDetect\footnote{\url{https://www.scenedetect.com/}} package to extract key frames representing different visual scenes corresponding to each sentence’s time interval. Next, we use GPT-4o to generate visual descriptions for each sentence by passing in all associated key frames. The model is prompted to describe the cooking steps, as well as the appearance, relative position, and relationships between ingredients and kitchenware. We use GAMA~\cite{ghosh-2024-gama} to generate text-based descriptions for environmental sounds such as sizzling or simmering. The extracted video information is compiled into a structured JSON file.

Before the system starts running, a user can configure the speed of text-to-speech according to their preference. During runtime, it automatically returns to the initial state (state 0) either when the user indicates satisfaction or when the system remains idle for a set period (5 seconds in our implementation), as shown in the state machine illustration in Fig.~\ref{fig: state_machine}.

\section{User Study}
To evaluate the usability and effectiveness of \toolname~ and to understand how BLV users use \toolname~ in cooking tasks, we conducted a study with eight BLV users\footnote{This study was approved by the IRB at our institution.}. The research questions are:
\begin{itemize}
    \item[RQ1:] How effective is \toolname~ for assisting BLV users in cooking. \looseness=-1
    \item[RQ2:] How do BLV users perceive the agency, control, and helpfulness when collaborating with \toolname~ in cooking?
    \item[RQ3:] What is the mental process of the user when coordinating with complementary perceptions from the system?
\end{itemize}

\subsection{Participants}
We recruited eight BLV participants for the study. Participants were screened using a demographic questionnaire to confirm that their visual acuity was worse than 20/200 and that they had no physical or medical conditions affecting mobility or the ability to handle kitchen tools. We did not put strict screening criteria on the prior cooking experience of potential participants. The average age of the participants was 37.1 ($\sigma = 13.7$). Five participants were female, and three were male. Detailed demographic information is shown in Table~\ref{tab: participants}.

\begin{table*}[!htb]
\centering
\begin{tabularx}{\textwidth}{c c c c X X X}
\toprule
\textbf{ID} & \textbf{Age} & \textbf{Gender} & \textbf{Onset} & \textbf{Level of Visual Impairment} & \textbf{Occupation} & \textbf{Cooking Frequency} \\
\midrule
P1 & 34 & M   & Congenital & Blindness with some light/color perception & Professor & About once a week \\
P2 & 33 & F & Congenital & Total Blindness & Self-employed & About once a week \\
P3 & 33 & M   & Congenital & Blindness with some light/color perception & Massage therapist & Almost every day \\
P4 & 25 & F & Congenital & Total Blindness & Student & Less than once a week \\
P5 & 62 & M   & Acquired   & Total Blindness & Chef & Almost every day \\
P6 & 28 & F & Congenital & Total blindness & Student & About once a week \\
P7 & 24 & F & Acquired   & Blindness with some light/color perception & Student & Less than once a week \\
P8 & 58 & F & Congenital & Total Blindness & Massage therapist & 2--4 times a week \\
\bottomrule
\end{tabularx}
\caption{Participant demographics for our user study}
\label{tab: participants}
\Description{A table summarizes participant demographics information for 8 participants (P1 to P8). For each participant, it lists age, gender, onset of blindness (congenital or acquired), level of visual impairment (either total blindness or blindness with some light/color perception), occupation (e.g., professor, student, chef), and typical cooking frequency (ranging from less than once a week to almost every day). Ages range from 24 to 62; five participants are female and three are male. Most participants have congenital blindness.}
\end{table*}

\subsection{Study Setting}

We conducted the user study in participants' own kitchens (P3, P5, and P8) or at alternative locations of their choice. We prepared three video recipes (details in Table~\ref{tab: video_user_study}), and the participant can freely choose one of them. Noticeably, we do not expect to draw any conclusions about the differences across video recipes, especially given the small sample size. These videos were preprocessed by \toolname~ using the knowledge extraction pipeline described in Sec.~\ref{sec:system}. All ingredients were prepared in advance by the experimenter and placed on a table. The participants were not aware of their exact positions; therefore, they used \textsc{Aroma} to explore and make confirmations. No heat-generating appliances or sharp knives were used during the sessions, per an Institutional Review Board (IRB) request to minimize the risk of the study protocol. In the three video recipes we selected, none of the steps require the use of sharp knives. However, some steps involve cutting tasks, such as dividing dough, that can be done using a safe knife. For any steps involving heat, participants were instructed to skip them. Each study session lasted about one hour, and participants were compensated with a 50 USD gift card for their time.

\begin{table}[!hb]
  \centering
  \begin{tabular}{@{}cllc@{}}
    \toprule
    \textbf{Index} & \textbf{Video ID}\footnotemark & \textbf{Recipe} & \textbf{Length} \\
    \midrule
    1  & lgg6luYfQ1w     & Pepperoni Pizza          & 6:18 \\
    2  & mixdagZ-fwI     & Beef Taco   & 5:45 \\
    3  & lH7pgsnyGrI     & Miso Soup            & 3:41 \\
    \bottomrule
  \end{tabular}
  \caption{Video recipes for user study.}
  \Description{A table listing the selected three video recipes for the user study. Namely, Pepperoni Pizza, Beef Taco, and Miso Soup}
  \label{tab: video_user_study}
\end{table}
\footnotetext{These videos can be accessed at \url{https://www.youtube.com/watch?v=[Video ID]}}

\subsection{Study Process}
After the consent process and a brief overview of the study procedure, the researcher informed the participant about the recipe categories listed in Table~\ref{tab: video_user_study}, and the participant selected one they liked from the list. The researcher then chose a different video recipe to use as an example for demonstrating the system's features. Specifically, each feature described in Sec.~\ref{sec: key_features} was demonstrated. During the process, the researcher also addressed any questions the participant had.

To begin the main cycle of the study, participants listened to the full video recipe to develop a high-level understanding of the dish, relying primarily on auditory cues. They were instructed to retain as much detail as possible to support their performance during the cooking session. 

Before the cooking process began, the researcher helped each participant wear a chest strap outfitted with a mobile phone that streamed a first-person video feed to the system. This chest-mounted setup was chosen based on prior findings showing that it offers greater comfort and more stable footage compared to head-mounted alternatives~\cite{li_contextual_2024}. The participant then attempts to cook the selected dish with the help of ~\toolname.

After completing the cooking task, participants filled out a post-study questionnaire, rating \toolname's usability, usefulness, and features on a 7-point Likert scale. Then, we conducted a semi-structured interview to understand participants’ experiences, perceptions, and expectations when using \toolname. The interviews lasted approximately 10–20 minutes and were audio-recorded for transcription and analysis. Specifically, we asked open-ended questions about participants’ overall impressions of the system, perceived usefulness of individual features, and how the interaction compared to their prior cooking experiences. We also followed up on specific moments we observed during the session, such as when participants made follow-up requests or deviated from the video instructions, to elicit their underlying reasoning and challenges. Additionally, for features rated highly or poorly in the questionnaire, we probed further to understand the rationale behind their ratings.

We conducted a thematic analysis~\cite{braun_using_2006} of the interview transcripts, recordings, and observational notes from the researcher. One researcher reviewed all the data and open-coded segments~\cite{corbin_grounded_1990} that reflected user behaviors, perceptions, and preferences. These initial codes were iteratively grouped into broader themes through axial coding, with regular discussion among the research team to refine interpretations and resolve discrepancies.

Additionally, we annotated the recording of each study session, labeling user requests, the events, and the system’s responses. We then evaluated \textsc{Aroma} on two aspects: (i) the accuracy of mapping each request to the corresponding user-initiated event shown in Fig.~\ref{fig: state_machine}. The accuracy is calculated by dividing the correctly mapped user queries by the total number of user queries, excluding user follow-ups (events 7 and 8). To do this, two researchers manually annotated the category of events for each request as the ground truth. Non-agreement cases were discussed to reconcile differences. (ii) the initial factual accuracy of the resulting response, without any user follow-ups (states 1-4 as shown in Fig.~\ref{fig: state_machine}). We compute it as the proportion of answers labeled correct out of all user queries. Similarly, the factual boolean labels are manually labeled by the researchers. Noticeably, we do not focus on the response quality in this analysis---this is reflected separately in the qualitative results from the user study. A response is considered correct if it meets both requirements: a.) it directly addresses the user's query; b.) the response contains no factual errors.\looseness=-1

\section{Results and Findings}
\begin{figure*}[!htb]
    \centering
    \includegraphics[width=0.9\linewidth]{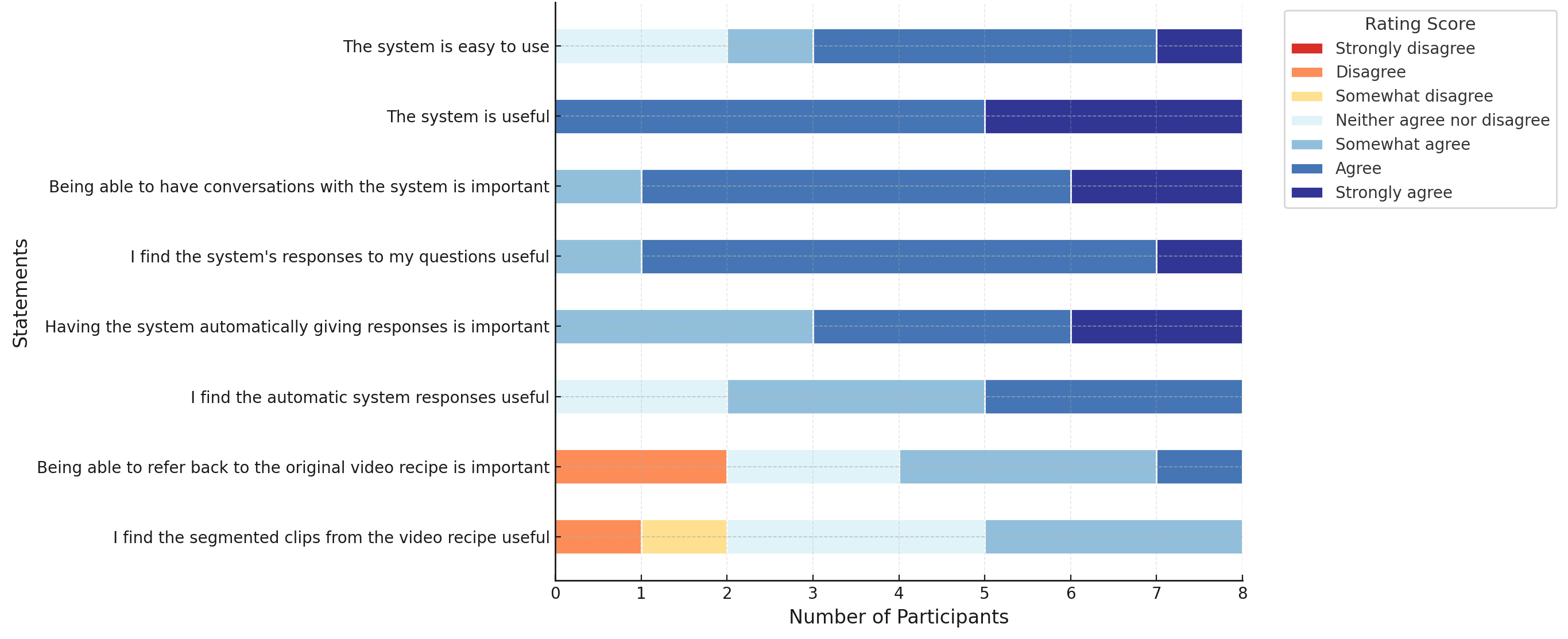}
    \caption{Participants’ ratings of the usability and usefulness statements of \toolname}
    \Description{A diagram showing the summary of user ratings to the usability questions. The distribution of user ratings for each usability question in AROMA shows that the highest frequency is for the ratings ``6'' and ``7'', indicating a positive attitude from the participants.}
    \label{fig: post_questionnaire}
\end{figure*}
In this section, we begin by presenting quantitative results on the system’s accuracy in mapping user requests to the correct events, as well as the accuracy of its immediate responses, which are generated in states 1–4, as shown in Fig.~\ref{fig: state_machine}. Then we report the overall usability ratings from participants, user feedback on each feature, and the reasons behind their perceptions. We then closely examine the specific requests made by BLV users. Building on this, we analyze their mental models when utilizing the additional perception and cognitive capabilities of \toolname.

\subsection{Quantitative Results on System Performance}
Participants frequently issued requests to \textsc{Aroma}. Table \ref{tab: quant_performance} reports the total request count, the system’s accuracy in mapping each request to the correct event in Fig.~\ref{fig: state_machine}, and the factual accuracy of its immediate responses. Across all participants, event-mapping accuracy averaged 0.82, while factual accuracy averaged 0.67.

Most mapping errors stemmed from the diverse ways participants phrased the same type of request. For example, when users sought clarification on a specific step (event 2), one might explicitly ask, \textit{``Could you explain this step for me?''} while another might phrase it more colloquially, e.g., \textit{``What's going on now?''}, which occasionally caused the model to misclassify the query as event 4, in which the backend LLM is prompted to give a generic response rather than step-specific guidance. Although such replies could be factually accurate, they did not satisfy users’ informational needs and undermined their experience. One of the participants reflected on this as: \textit{``When I ask about a step, I expect details like how long it takes and how to perform it, but I remember once the system just simply told me what this step was.''} As \textsc{Aroma} relies on prompts for event mapping, we anticipate that incorporating user-specific few-shot examples in the future will reduce these errors and enhance user experience.

\begin{table}[!b]
  \footnotesize                      
  \centering
  \begin{tabular}{@{}c c c c c c@{}}
    \toprule
    \textbf{Part.} &
    \makecell{\textbf{Total}\\\textbf{Queries}} &
    \makecell{\textbf{Correct}\\\textbf{Mappings}} &
    \makecell{\textbf{Mapping}\\\textbf{Accuracy}} &
    \makecell{\textbf{Correct}\\\textbf{Responses}} &
    \makecell{\textbf{Response}\\\textbf{Accuracy}} \\
    \midrule
    P1 & 10 & 9  & 0.90 & 8 & 0.80 \\
    P2 & 6  & 5  & 0.83 & 5 & 0.83 \\
    P3 & 13 & 11 & 0.85 & 9 & 0.69 \\
    P4 & 7  & 5  & 0.71 & 4 & 0.57 \\
    P5 & 16 & 12 & 0.75 & 10 & 0.63 \\
    P6 & 11 & 9  & 0.82 & 8 & 0.73 \\
    P7 & 8  & 7  & 0.88 & 5 & 0.63 \\
    P8 & 12 & 10 & 0.83 & 7 & 0.58 \\
    \bottomrule
  \end{tabular}
  \caption{System performance across participants}
  \label{tab: quant_performance}
  \Description{A table that reports quantitative results for 8 study participants (P1–P8). For each participant, the table shows: total number of user queries, number of correctly mapped queries to system events, mapping accuracy, number of correct system responses, and response accuracy. Mapping accuracy ranges from 0.71 to 0.90, while response accuracy ranges from 0.57 to 0.83.}
\end{table}

Response errors most often arise from hallucinations in the underlying language model. As explained in Sec.~\ref{sec:system}, \textsc{Aroma} answers each user-initiated query by conditioning on the runtime session history. As this context window lengthens, the accuracy degrades accordingly, which is a critical limitation of most LLM-based systems~\cite{liu2024lost,lu2024controlled}. To mitigate this problem, especially for future deployments that support BLV users tackling more complex recipes, the underlying algorithm of the context (memory) managing module (Fig.~\ref{fig: system_arch}~{\textcolor[HTML]{EE754D}{\rule{0.6em}{0.6em}}}) should be enhanced to reduce the redundancy of the memory, and enable more efficient searching algorithm for retrieving the appropriate context, rather than purely relying on the LLM capability.

\subsection{System Usability and User Perception}
Overall, participants reported high satisfaction with \toolname’s usability and its effectiveness in supporting cooking tasks. Post-study questionnaire results indicate that conversational features, including receiving immediate responses to questions and asking follow-ups (Sec.~\ref{sec: ft_user_init}), were particularly well-received. These features were rated as both highly important ($\mu = 6.13$, $\sigma = 0.60$) and useful ($\mu = 6.00$, $\sigma = 0.50$). System-initiated assistance (Sec.~\ref{sec: ft_proactively_monitoring}) also received strong importance ($\mu = 5.88$, $\sigma = 0.78$) and ($\mu = 5.13$, $\sigma = 0.78$) useful ratings, suggesting that participants appreciated the system's ability to monitor progress and offer timely support without prompting. Lastly, the ability to replay segments of the recipe video (Sec.~\ref{sec: ft_access_offline}) was not seen as important as the previous ones ($\mu = 4.13$, $\sigma = 1.36$), reflecting its role as a supplementary aid rather than a critical component of interaction. Fig.~\ref{fig: post_questionnaire} summarizes the distribution of participant ratings across these features.

\subsubsection{Conversational interaction supports flexible access to video recipes, enhances agency, and reduces cognitive load for the user}
Participants emphasized that conversational interaction allowed them to engage with content more flexibly and on their own terms, rather than being constrained by the linear structure of a video. For instance, P3 appreciated being able to re-ask questions and request clarifications: \textit{``I could ask what ingredients I need for the next step without hearing back from the video again.''} This ability to access information non-linearly helped reduce stress, especially when compared to traditional video consumption. As P1 noted: \textit{``With a video, once you miss something, it's hard to go back and catch it.''} \looseness=-1

Going beyond, when system outputs were ambiguous or incomplete, users often followed up with contextual cues based on their own non-visual perception to refine or confirm the answer. For example, P5 first asked the agent for an overview of the kitchen setup, then used that as a scaffold to ask specifically about the location of the pizza dough: \textit{``This helps me build a hierarchical understanding of what’s on my counter.''} Similarly, P4 recounted using touch to distinguish between two plates when the system ambiguously referred to ``a round plate'': \textit{``I could feel there were two plates. So I picked one (from the texture) and asked, `Is this the pepperoni?' and got confirmation. That helps me be sure.''}

\subsubsection{Proactive feedback was helpful when precise and well-timed}
Several participants found that proactive feedback helped them stay on track and feel more confident moving forward. For example, P6 recalled, \textit{``The system automatically detected a wrong step where I was about to put the miso paste before adding the tofu into the soup.''} Similarly, P2 and P3 appreciated the reminders to roll the pizza crust edge to cover the string cheese---something they had missed, but the system intervened with a useful reminder.. These interventions reduced their cognitive load and stress, as P3 noted: \textit{``I know I have an assistant by my side to detect some, if not all, problems.''}

The challenge of this feature, however, was the timing and relevance of the interventions. Participants were frustrated when the system intervened during intentional deviations. For instance, P5 skipped a cutting step due to difficulty, yet the system still prompted her to complete it. Additionally, due to inevitable time delays, responses sometimes failed to align precisely with user actions. Despite these issues, precise proactive messages were generally well-received, as they could be easily ignored. As P2 described, the system felt like a ``background assistant'' rather than a distraction.

\subsubsection{Segmented video replays were useful when original video clips were accessible}
In general, participants found the feature useful when the original audio information---instructions from the speaker, for instance---was accessible and informative. In these cases, replays helped users quickly \textit{``verify step sequences''} (P6) or confirm specific actions without explicitly asking the system to answer it, which takes extra time (P2). However, the usefulness of this feature is limited for inaccessible video clips. For example, P3 noticed that: \textit{``I tried this feature several times, but then I realized I couldn't get enough information I wanted, compared with other features.''}

\subsection{What Requests are BLV Users Making?}
To better understand how BLV participants interacted with \toolname\ over cooking, we analyzed the types of requests they made throughout the sessions. Drawing from both observation and interview data, we identified key categories of requests, their underlying motivations, and implications for future system design.

\subsubsection{Maintain awareness of procedural flow}
In our study, participants frequently relied on the system for procedural guidance. Nearly all participants asked questions such as \textit{``What should I do next?''} (P1) and \textit{``I have finished preparing the meat, what’s the next step?''} (P6). These requests were especially common when participants had already made several other requests in the previous step, or when they wanted to skip a specific step---for example, a participant asked \textit{``What should I do after baking the pizza?''} (P2). Addressing this, future systems could consider automatically responding when detecting the completion of a step. However, determining the ``end of a procedure'' is inherently ambiguous and may place greater demands on the model’s performance and accuracy.

\subsubsection{Confirm the types and locations of foods and ingredients}
In addition to procedural support, participants often used the system to confirm the identity or location of ingredients, especially when non-visual sensory cues alone were insufficient.  Specifically, participants often turned to the AI system when multiple items had similar textures, smells, or spatial positions. For example, P4 asked questions like \textit{``Is this the mozzarella?''} when she felt a moist texture, but was not sure whether it was mozzarella or pepperoni. In other cases, users directly sought help from the visual agent to locate items. For instance, P5 asked, \textit{``Do you see anything that looks like pizza dough?''} These interactions illustrate a collaborative dynamic between user and system: participants brought non-visual sensory impressions to the conversation, while the AI contributed visual grounding. Together, they formed a hybrid perception loop, helping users confidently identify and locate items in their environment.

\subsubsection{Clarification and response customization through follow-up requests}
Participants frequently followed up on initial agent responses to clarify ambiguous instructions or to tailor answers to their individual preferences. For instance, P3 asked the system to \textit{``order it from top left to bottom right''} after asking \textit{``tell me what you see...''} as first, explicitly requesting a spatially structured response aligned to his habits. Similarly, P1 asked to \textit{``Describe what you see in a clockwise direction.''} and P4 asked to \textit{``Describe the items to my left one by one.''}

In other cases, users rephrased or elaborated their questions when the initial answer lacked sufficient detail or clarity. For example, P5 attempted to form a stuffed pizza crust and asked repeatedly about the 90-degree folding step, iterating: \textit{``Is this one the video said to rotate 90 degrees? Or am I doing it wrong?''}. These interactions emphasize the importance of supporting iterative, dialogic clarification---allowing users to gradually refine their understanding and guide the system toward more useful, personalized responses.

\subsection{How Non-Visual and Visual Perception and Cognition Complement Each Other?}
Our study revealed how BLV users integrate visual assistance from the AI system into their existing non-visual cooking strategies. Rather than replacing their tactile, auditory, olfactory, and gustatory skills, \toolname~ acted as a complementary scaffold that enhanced these well-practiced abilities. We observed several key patterns in how users combined visual and non-visual information during the cooking process:

\subsubsection{Non-visual perception as the foundation for assessing food state}
For participants who had experience in cooking, they began with a robust, non-visual schema for gauging food readiness and identifying ingredients. These strategies included interpreting texture, shape, container types, and spatial orientation through touch.  Although participants who had less cooking experience (P4, P8) also expressed that the additional visual information is \textit{``additional help''} (P8). Specifically, P5, who is an experienced blind chief, described how the use of various bowl shapes helps them identify ingredients without needing to touch the ingredients directly: \textit{``I use different shapes and sizes and textures. So if the computer couldn’t answer me, I could.''} Similarly, P3 explained how he could \textit{``feel the edge of the pan so I know how far I need to stretch the dough.''} In another instance, a participant confirmed a sauce’s identity by taste rather than asking the agent. 

These examples illustrate that many users used the AI primarily to validate their own perceptions. As P5 reflected: \textit{``You do know by touch that the crust is uniform without asking the computer''}.

\subsubsection{Resolving ambiguities in object and step identification.} 
Participants frequently used \toolname~ to verify what had inferred through touch or sound, especially when encountering ambiguous objects. As P5 explained while handling pizza dough: \textit{``I think this is probably triangular... I can tell by touch, but I’m just confirming.''}. This interplay of tactile inference and visual confirmation was especially important when participants encountered objects with similar textures or forms. For instance, P2 described struggling to distinguish between string cheese and mozzarella by touch or taste alone, noting that \textit{``they all have similar textures and flavors, and the video doesn’t say what they look like either.''} In another moment, when the participant (P1) did not understand how to shape the crust with chopped cheese, they explicitly asked: \textit{``Did the video say anything about rotating 90 degrees?''} and later double-checked by asking: \textit{``Is this what the video said about folding it?''}, showing their reliance on the visual properties of the food at this moment.

\subsubsection{Improving spatial awareness of ingredients and tools.} 
Participants used the AI's egocentric vision not only to confirm objects' identities but also to orient themselves within the physical workspace. For example, during the preparation of each dish, nearly every participant asked the system to help locate ingredients and cookware, and confirm if the ingredient is in the correct position. For instance, P2 asked questions such as: \textit{``Am I spreading the pizza sauce on the center of the dough?''}. We also observed that participants combined their demonstrations with the system’s visual feedback to ask spatial questions more directly, such as: \textit{``If cheese sticks were evenly placed here?''} (P3)---while simultaneously pointing to the pizza crust with their finger. 

\subsubsection{User expectations: extending visual perception beyond food and ingredients to include gestures from users}
In \toolname~, the real-time visual analysis agent was originally designed to visually track and interpret the ingredients and tools in the process. However, from the creative uses of the participants during the study, we observed opportunities to improve their experience. Specifically, P6 once requested to use schemas like ``clockwise direction'' for the system to describe objects. Meanwhile, as noticed by P3, he prefers responses based on hand-relative coordinates (e.g., “left of your right hand”), while others like P4 don't show a particular preference for this.

These observations suggest two promising directions for future versions of the system. First, responses could be personalized based on users’ preferred spatial language or cognitive mapping strategies. Second, incorporating gesture recognition could provide additional context, improving the accuracy and relevance of visual feedback in real-time interactions.

\section{Discussion}

Our study of \toolname\ highlights that designing assistive technologies for BLV users requires more than providing additional sensory information — it demands systems that collaborate with users' embodied skills, adaptive strategies, and personal expertise. We discuss how future systems can move toward human-AI co-reasoning, support multimodal and sensor-integrated interactions, and balance proactive assistance with user privacy and agency. Finally, we reflect on how \toolname's design principles may generalize to other activities of daily living (ADLs), informing more flexible, inclusive, and context-aware assistive technologies.

\subsection{From Providing Additional Sensory Information to Co-Reasoning}
Our study shows that BLV users engage in cooking not just by the additional visual information, but through a rich, embodied non-visual cognitive ability developed over time, which is grounded in touch, smell, sound perception, and their spatial memory etc. Therefore, we suggest that assistive technology should go beyond simply asking, \textit{``how to compensate what is missing?''}, future systems should consider \textit{``how to work with the user's existing and ongoing cognitive process?''} Just as P5 explained: \textit{``I use different shapes and sizes and textures. So if the computer couldn’t answer me, I could.''}

In \toolname, we leverage the text-based reasoning capability of LLMs to contextualize its response based on non-visual clues given by the user through prompting. However, it is inevitable that the responses will occasionally be incorrect, ambiguous, or not actionable (for instance, providing instructions that have visual descriptions) --- depending on the performance of the model itself. In \toolname\, we try to reduce such effects by carefully designing the prompts that we use, looking ahead, future systems could build on this foundation by incorporating specialized models that offer greater robustness and reliability in interpreting and responding in non-visual contexts.

\subsection{Multimodal User Interaction Beyond Verbalization}
\toolname's current interaction paradigm primarily leverages spoken commands and perceptual descriptions provided by the user. However, in the user study, we also noticed that participants naturally employed a variety of non-verbal cues for expression throughout their cooking processes. For instance, we observed frequent gestures such as pointing, tapping, or sweeping hand movements to reference objects or to indicate spatial queries. This observation is consistent with prior research that emphasizes the potential of hand- or wrist-based gestures for supporting non-visual interaction, such as navigating to the next step in a recipe \cite{li_non-visual_2021}. A potential improvement could be integrating a robust gesture-recognition component to interpret these signals \cite{li_non-visual_2021}, which would augment the visual understanding capability of the system beyond a simple end-to-end LLM call and further reduce the cognitive load of verbalizing for BLV users.

Going beyond gestures that have explicit semantic meaning, tracking user expressions and micro-gestures could also serve as effective cognitive indicators~\cite{haggard_micromomentary_1966}. Specifically, indicators such as hesitation pauses, head movement trajectory could all serve as indicators of cognitive load, confusion, or uncertainty. Integrating those non-verbal cues in the memory management component of \toolname\ (see Fig.~\ref{fig: system_arch} {\textcolor[HTML]{EE754D}{\rule{0.6em}{0.6em}}}) and adapting the context dispatch strategy accordingly, presents an opportunity for \toolname\ to deliver results that are better contextualized to the user's cognitive state.

\subsection{Toward Multimodal, Sensor-Integrated Interaction Paradigms}
In the current design of \toolname, users verbalize their sensory experiences, offering cues beyond visual information to help the system determine the cooking state and provide appropriate guidance for next steps. However, several participants noted that this process can be cognitively demanding.

Sensor augmentation offers a path forward: thermal sensors, smart utensils, visual object/action recognition~\cite{li_oscar_2025}, or even audio classifiers could reduce this burden by allowing the system to proactively support users based on real-time sensor signals. These approaches could support a more seamless mixed-initiative flow, where AI agents mediate the timing and content of intervention based on contextual uncertainty and user need.

Yet, our study reaffirms that embodied, subjective knowledge, such as taste or smell, remains essential to cooking. Even in sensor-rich environments, systems should preserve and support this form of expertise. This also aligns with Dourish’s view of embodied interaction~\cite{dourish_where_2001}, where technology is most effective when it is grounded in the body and lived practice. \toolname\ demonstrates that subjective perception is not only valid input, but often the most reliable and personally meaningful cue in practice.

\subsection{Designing for Procedural Fluidity vs. Procedural Fidelity}

While \toolname\ was designed to support users in following the procedural flow of video recipes, our findings reveal that BLV users often engage in cooking practices that intentionally diverge from strict procedural fidelity. Participants might skip, reorder, or modify recipe steps to accommodate personal preferences, available ingredients, or contextual constraints. These observations highlight an important future direction for assistive systems: designing for procedural fluidity rather than enforcing rigid adherence to predefined steps. Future systems could model users' intent to distinguish between errors and deliberate adaptations, enabling more collaborative interaction \cite{li_non-visual_2021}. For instance, when users skip or combine steps for efficiency or convenience, the system could proactively adjust its guidance to reflect the new procedural context. This design approach respects the autonomy and expertise of BLV users while maintaining safety and alignment with instructional content.

\subsection{Balancing Proactivity with Privacy and Agency in Sensor-Rich Environments}

\toolname's mixed-initiative design, particularly its proactive monitoring features, was generally well-received; however, participants' experiences also underscore the importance of balancing system proactivity with user agency and privacy considerations. While timely interventions helped prevent errors and supported task progression, participants occasionally expressed frustration when the system intervened during intentional deviations or when proactive prompts conflicted with user preferences. This highlights an emerging design challenge for future assistive technologies operating in personal or sensor-rich environments \cite{ahmed2015privacy}. Systems should enable configurable levels of proactivity, allowing users to specify preferences for monitoring sensitivity, intervention types, or contexts in which proactive assistance is appropriate. Designing consent-driven and context-aware proactivity mechanisms would preserve user control while offering tailored support aligned with individual privacy expectations and situational needs.

\subsection{Extending \toolname's Framework to Other Activities of Daily Living}
While our work focuses on cooking, the design principles underlying \toolname---grounding real-world sensory perception with instructional video content through mixed-initiative interaction---hold promise for supporting other activities of daily living (ADLs). Tasks such as makeup application~\cite{li2022feels}, home repair \cite{leporini2020designing}, arts and crafts \cite{li2023understanding}, or gardening \cite{lambe1995gardening} similarly involve multimodal perception, procedural knowledge, and embodied expertise. Each domain presents unique sensory demands and interaction challenges, necessitating domain-specific adaptations. Nevertheless, the core approach of leveraging user-provided non-visual cues, integrating real-time visual analysis, and aligning with video-based instruction could be generalized across ADLs. Future research should explore the transferability of \toolname's framework, contributing to the development of universal assistive technologies that support BLV users across diverse everyday activities.
\section{Limitations}
While our system demonstrates the potential of AI-enhanced, non-visual cooking support through multimodal grounding with video recipes, the current implementation reveals several limitations that suggest important directions for future work.

Firstly, our user study involved eight BLV participants, which is a limited sample size. In addition, heat and sharp knives are not used due to safety concerns. Therefore, the study may not fully represent the range and steps of cooking activities that BLV users routinely perform. In contrast, \textsc{Aroma} is a proof of concept that shows how a human–AI partnership can combine visual and non-visual cues to perceive the kitchen state and translate video-based knowledge into actions in reality.

Another limitation is the unstable latency and response timing of the backend multi-modal LLMs that affect the user experience. Despite leveraging the Gemini multimodal API and OpenAI's real-time audio transcription to minimize delay, response latency remains a significant limitation in both system-initiated and user-initiated interactions. For system-initiated prompts, such as real-time visual analysis of the user’s current cooking state, visual processing occasionally lagged behind action. This led to situations where \textit{``The agent confirmed a step I already did 5 seconds ago.''} as P7 noted. This delay may also compromise safety in time-sensitive steps. User-initiated queries also experienced varying response times depending on request complexity and length. For example, when a user asked, \textit{``Can you tell me what you see right now? And order it from top-left to bottom-right?''} (P4), the response required full-scene analysis, leading to a multi-second pause that broke task flow. Participants expressed frustration with these lags in the interview: \textit{``Sometimes the response is just not perfect, or I should say it’s too slow''} (P3). Improving model optimization and pipelining inference tasks in parallel is crucial to achieving more fluid and dependable interaction.

In addition, the current version of \toolname~ requires preprocessing the video offline, which may limit recipe access for BLV people. The instructional video must first be parsed to extract stepwise procedural knowledge, ingredient states, and visual targets --- a process we have not implemented on-the-fly. In the interview, we were frequently asked questions like \textit{``What if I want to try something from another YouTube channel? Can it still help me?''} This motivates future work on scalable pipelines that can process arbitrary video recipes near real-time to support ``plug-and-play'' recipe access.

\section{Conclusion}
In this work, we present \toolname, a mixed-initiative, multimodal AI system that supports BLV users in cooking by grounding their real-time non-visual sensory input and first-person video stream from a wearable camera with the audiovisual information from a video recipe. Through a study with eight BLV participants, we demonstrate the effectiveness of \toolname~ and highlight the insights for designing AI systems that not only provide additional visual information but also recognize and respond to the unique perceptual states of BLV users in performing real-world tasks. 

\begin{acks}
This work was supported in part by a Notre Dame-IBM Technology Ethics Lab Award, an NVIDIA Academic Hardware Grant, a Google Research Scholar Award, a Gift from Adobe Inc., the National Eye Institute of the National Institutes of Health R01EY037100, and NSF CMMI-2326378. Any opinions, findings, or recommendations expressed here are those of the authors and do not necessarily reflect the views of the sponsors.
\end{acks}

\balance
\bibliographystyle{ACM-Reference-Format}
\bibliography{references,references-2}

\clearpage
\onecolumn

\appendix
\end{document}